\documentclass[12pt]{iopart}
\usepackage{graphicx} 
\usepackage[utf8]{inputenc}
\usepackage{caption}
\usepackage{subcaption}
\usepackage{xcolor}

\usepackage[numbers]{natbib}
\usepackage{notoccite}
\def\newblock{\ }

\newcommand{\bQ}{\mathbf{Q}}
\newcommand{\bk}{\mathbf{k}}

\begin{document}

\title[]{Tuning the non-linear interactions of hybrid interlayer excitons in bilayer MoS$_2$ via electric fields}
\author{Mathias Federolf}
\address{Lehrstuhl für Technische Physik, University Würzburg}
\author{Alexander Steinhoff}
\address{Institut für Physik, Carl von Ossietzky Universität Oldenburg}
\author{Monika Emmerling}
\address{Lehrstuhl für Technische Physik, University Würzburg}
\author{Matthias Florian}
\address{Department of Electrical Engineering and Computer Science, University of Michigan}
\author{Simon Betzold}
\address{Lehrstuhl für Technische Physik, University Würzburg}
\author{Christian Schneider}
\address{Institut für Physik, Carl von Ossietzky Universität Oldenburg}
\author{Sven H\"ofling}
\address{Lehrstuhl für Technische Physik, University Würzburg},

\ead{mathias.federolf@uni-wuerzburg.de}

\begin{abstract}

Hybrid interlayer excitons in bilayer MoS$_2$ are a promising platform for nonlinear optics due to their intrinsic dipolar character, which combines in-plane and out-of-plane dipole moments. 
In this work, we directly probe the nonlinear exciton-exciton interactions of hybrid interlayer excitons. By applying an external out-of-plane electric field, we polarize the excitons to enhance their mutual dipolar interactions, thereby deliberately favoring these repulsive contributions over competing attractive many-body corrections.
We furthermore establish a fully microscopic theoretical description of these effects to explain the core experimental results. The tuning results in a significantly larger blueshift compared to the zero-field case and perspectively opens an avenue to even switch between repulsive and attractive interaction potentials. Our findings establish that strong nonlinearities can be tuned via an external electric field, providing a new degree of control over exciton interactions beyond density tuning alone.

\end{abstract}

\section{Introduction}
\label{sec:introduction}

Interlayer excitons in van der Waals heterostructures have attracted great interest due to their enhanced dipolar interactions \cite{rivera_interlayer_2018} and non-linearities \cite{steinhoff2024exciton}, enabling the formation of complex excitonic quantum phases \cite{moody_exciton_2016}. However, due to their relatively low oscillator strength, they play a minor role in exciton-polariton research. In turn, in homobilayer MoS$_2$, interlayer excitons can couple to intralayer excitons, forming hybrid interlayer excitons (hIEs). In this case, an electron in one layer interacts with a hole tunneling between both layers \cite{gerber2019interlayer}. This leads to excitons with both out-of-plane and in-plane dipole moments, which enhances their oscillator strength \cite{lorchat2021excitons}. An external electric field lifts the degeneracy of the dipole orientation, splitting the hybrid interlayer exciton into two distinct states corresponding to dipole moments parallel and antiparallel to the applied field \cite{leisgang2020giant, lorchat2021excitons, peimyoo2021electrical}. 

So far, strong nonlinearities of hIEs have primarily been investigated in the exciton--polariton regime, where coupling to a cavity mode alters the interaction effects \cite{datta2022highly, louca2023interspecies}. However, the  nonlinearity of the hybrid interlayer exciton states under an external electric field in bilayer MoS$_2$ has not been explored. In this work, we demonstrate that nonlinearities can be controlled not only by excitation density but also with external fields. By applying an out-of-plane electric field, we show that the polarization of hIEs leads to enhanced exciton-exciton repulsion and a significantly stronger blueshift compared to the zero-field case. 
Our experimental findings are supported by an excitonic many-body theory based on first-principle band structures and Coulomb interaction matrix elements along the lines of Ref.~\cite{steinhoff2024exciton}. Besides the classical dipole-dipole interaction, the approach explicitly treats exchange effects due to the fermionic substructure of excitons as well as dynamical screening effects to compute density-induced energy renormalizations. Our calculations reveal that due to compensation between various attractive and repulsive contributions, the net exciton energy shifts are very sensitive to the selective tuning of individual interaction terms.
This establishes bilayer MoS$_2$ as a robust platform where strong excitonic nonlinearities can be directly observed and controlled, providing the groundwork for tunable polaritonic implementations. 

\section{Results}
\label{sec:results}

\begin{figure}
     \centering
     
     \includegraphics[width=\textwidth]{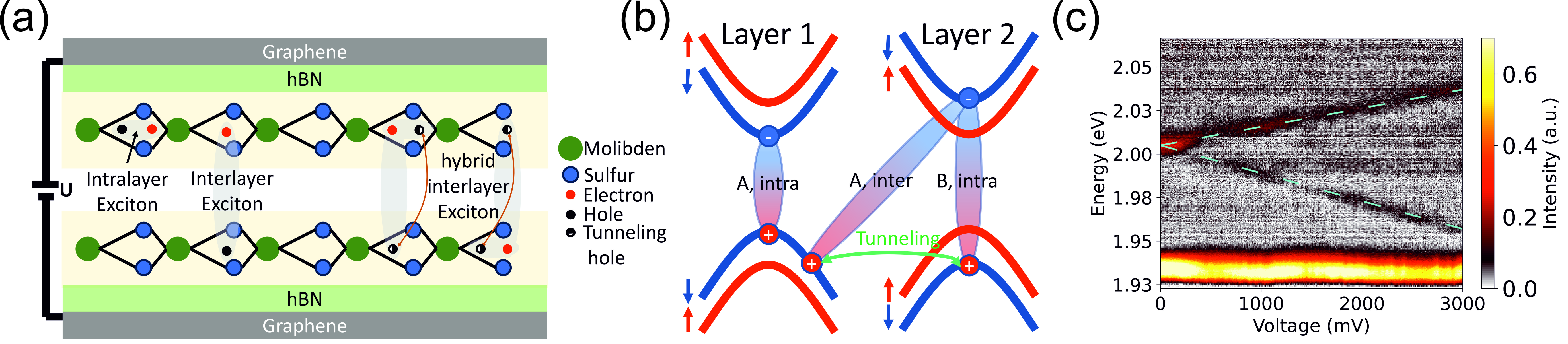}
          
        \caption{(a) Real-space structure of the emergent exctions in bilayer MoS$_2$, composed of intra- as well as hybrid interlayer excitons.  (b) Band structure illustration of the A interlayer exciton and the hIE around the K point. The tunneling of the hole between the two layers is indicated by the green arrow. In (c) we vary a voltage across the two graphene layers and visualize the splitting of the hIX by measuring the reflection contrast.}
        \label{fig:skizze}
\end{figure}

Hybrid interlayer excitons are a special class of charge complexes in 2D systems as they exhibit both in-plane and out-of-plane dipole moments. In these systems the A interlayer exciton couples to the B intralayer exciton (see sketch in Figure \ref{fig:skizze} (a)). The corresponding momentum-energy diagram is shown in Figure \ref{fig:skizze} (b). Our sample consists of a stack of graphene / hBN / bilayer MoS$_2$ / hBN / graphene, schematically shown in Figure \ref{fig:skizze} (a). The gold contacts were etched into the SiO$_2$ cover layer on the wafer before the flakes were placed on top. Stacking was performed using the viscous-elastic stacking method \cite{castellanos-gomez_deterministic_2014}. 
We probe the optical response of our field-tunable bilayer sample via white-light reflection measurments at temperatures of 4K. The reflection contrast is calculated as $R_{back}-R_{data}/R_{back}= \Delta R/R$, with $R_{back}$ beeing the background signal recorded on a reference on the gold contact.
We first confirm the general field-tunability of the hybrid exciton by ramping the voltage between the two graphene contacts in the capacitor-like structure.
We apply a voltage using the two graphene flakes in the capacitor-like structure. This results in a splitting between the hlX in the parallel and antiparallel state (in reference to the electric field vector), which clearly exhibits its fingerprints in the reflection contrast in Fig 1c). In stark contrast, the reflection signal of the A exciton remains vastly unaffected by the external field. For only 3V the splitting between the two hlX complexes is as large as 81 meV. 

\begin{figure}
     \centering
     \includegraphics[width=\textwidth]{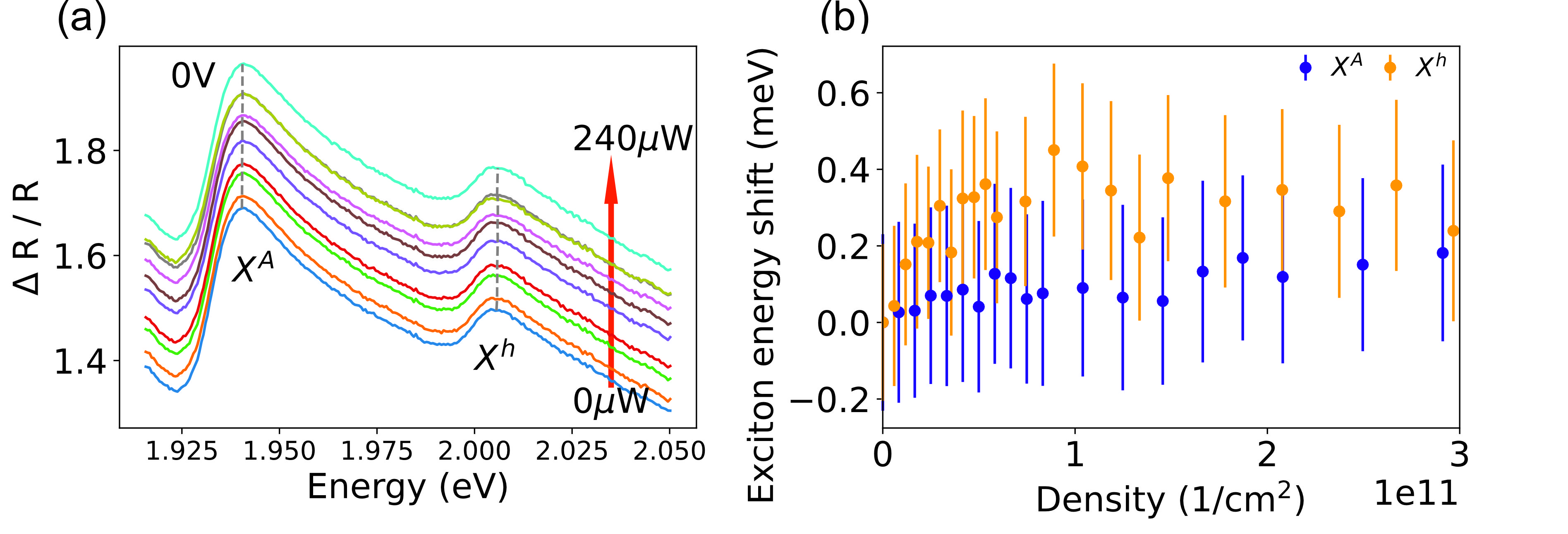}
        \caption{(a) Reflectivity measurements of bilayer MoS$_2$ at zero gate voltage. Increasing the power of an additional green laser from 0 $\mu$W to 240 $\mu$W results in a net blue shift of the two exciton peaks. The peak around 1.94 eV corresponds to the intra-layer A exciton, while the peak at 2.01 eV is identified as the hIE. (b) Extracted energy shifts of the A exciton and the hIE exciton versus the density of the respective exciton species. 
        }
        \label{fig:0V}
\end{figure}

The optical non-linear response is next probed in density dependent studies. Therein, we utilize a green CW laser , which we focus on the same sample spot as the white light. In the zero-voltage case, the signal shows two distinct peaks: the A exciton around 1.935 eV and the hIE around 2.01 eV, see Figure \ref{fig:0V} (a).   
To measure the excitation-dependent behavior of the excitons,
the quasi-particle density was varied by using different powers of the green laser. The excitation energy of the green laser was varied, and at each step we recorded ten measurements, which were averaged to reduce detector noise due to the short integration times. 
The evaluated data are plotted in Figure \ref{fig:0V} (a), where the lowest excitation power corresponds to the lowest trace, with increasing green-laser power for higher traces. 
For each measurement the peak position was extracted by peak fitting using a model consisting of a decaying exponential background with two Gaussian peaks. The exciton densities were estimated using a rate equation model that takes into account the laser-induced creation of excitations above the gap as well as radiative recombination of various exciton species according to their respective oscillator strength. The overall exciton distribution is then deduced from the measured photoluminescence intensity, see the detailed discussion in the Appendix. 
Figure \ref{fig:0V} (b) shows the nonlinearity of both exciton species by plotting the fitted peak energy versus the calculated densities of the respective species. The hIE exhibits a slightly stronger nonlinearity than the A exciton, in line with Ref.~\cite{louca2023interspecies}. 

\begin{figure}[!h]
	\centering
     \includegraphics[width=\textwidth]{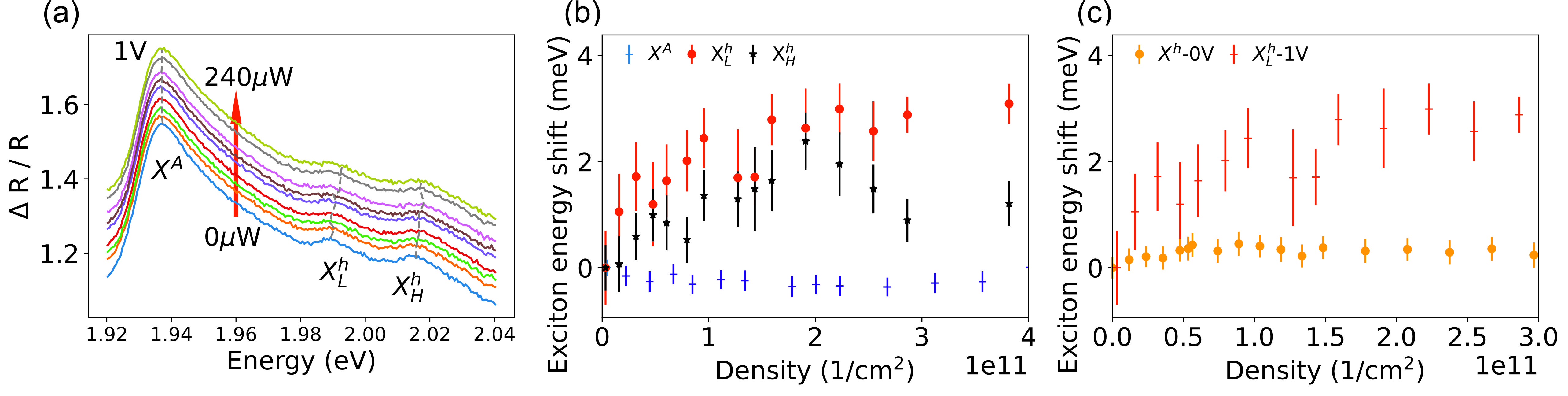}

     
        \caption{(a) Reflectivity contrast measurements at 1 V gate voltage, corresponding to a local electric field at the bilayer of roughly 0.1 V/nm. The A exciton peak is at 1.94 eV, while the hIE splits into a lower-energy state at 1.99 eV and a higher-energy state at 2.01 eV. (b) Extracted energy shifts of the A-exciton, the lower and the upper hybrid interlayer exction as a function of the respective densities, showing enhanced nonlinearities in the presence of the external electric field. (c) Comparison of blueshifts of hybrid interlayer exction for the 0V and 1V cases. The hIE state under finite electric field shows a significantly stronger density dependence than in the degenerate case at 0V. }
        \label{fig:1V}
\end{figure}

The zero electric field inter-layer dipole moments of the hIE are degenerate, as they can be oriented either up or down, in accordance with the bilayer inversion symmetry. This degeneracy can be lifted by applying an electric field that breaks the inversion symmetry, yielding a peak splitting (Fig. \ref{fig:skizze} (a)) \cite{lorchat2021excitons, leisgang2020giant, peimyoo2021electrical}. 
In Figure \ref{fig:1V} (a), we show such combined measurements performed under a gate voltage of 1 V, following the same procedure as in the zero-voltage case. 
Three peaks are now visible: the A exciton, the low-energy hIE, and the high-energy hIE, where the latter two result from a splitting of the hIE according to their (anti-)alignment with the external field.
The splittingunder the chosen experimental condition amounts to about 20 meV, from which one can deduce a local electric field at the bilayer of roughly $0.1$ Vnm$^{-1}$ \cite{peimyoo2021electrical}. 
Figure \ref{fig:1V} (b) shows the extracted peak positions versus the corresponding exciton densities obtained from the rate equation model. 
Particularly the lower hIE exhibits a significantly stronger blueshift than the A exciton, which directly reflects its dipolar nature, yielding enhanced interactions.
This becomes even more evident in Figure \ref{fig:1V} (c), where we directly compare the results for both gate conditions. The hIEs under a $0.1$ Vnm$^{-1}$ field exhibit a larger density-dependent shift than in the degenerate, zero-field case, confirming that field-induced dipole alignment enhances repulsive exciton-exciton interactions.

To gain further insight into the interactions in the homobilayer system, we calculate density-dependent exciton energy renormalizations of the A exciton and the hIE, expanding on the work discussed in \cite{steinhoff2024exciton}. Key to our method is the combination of  state-of-the-art many-body theory with material-realistic band structure and Coulomb matrix element calculations. Details on the band structures and Coulomb matrix elements are given in the Appendix. 
\begin{figure}[!h]
     \centering

         \includegraphics[width=\textwidth]{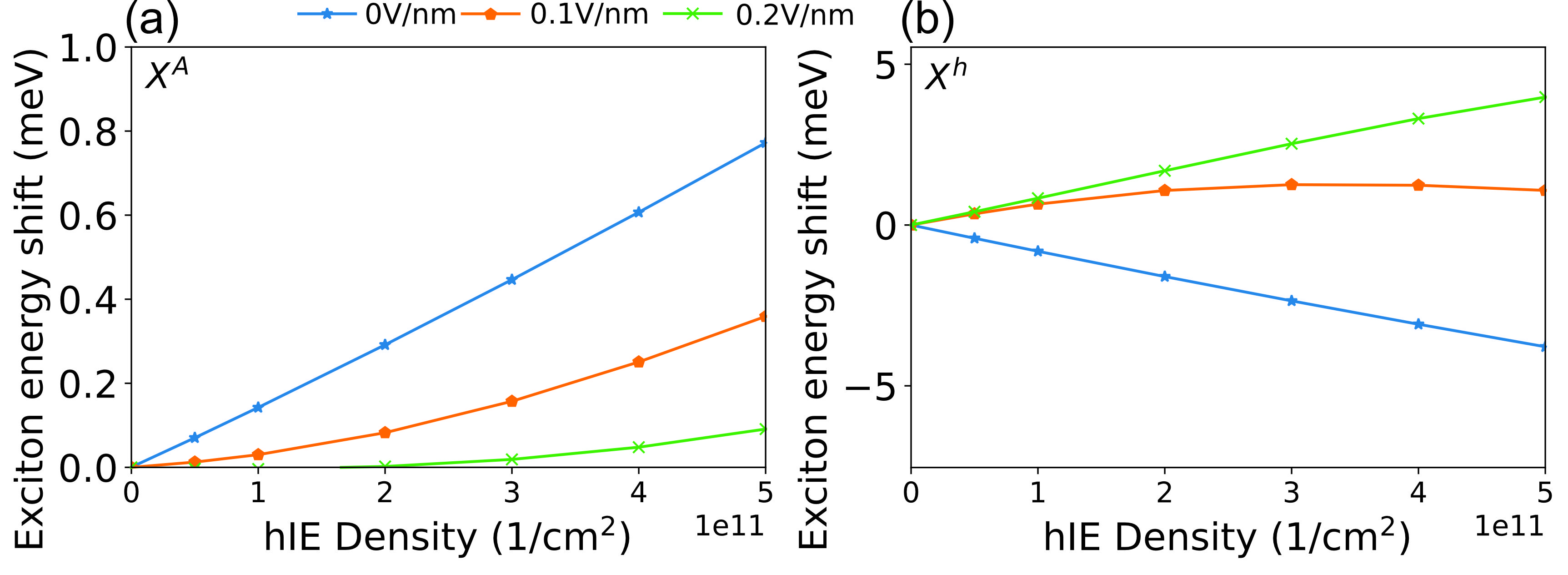}

        \caption{Calculated exciton energy renormalization induced by interaction with a reservoir of hIE for the A exciton (a) and for the lower hIE (b). The effective exciton temperature is $70$ K. Multiple voltages are shown to visualize the trend with increasing external electric field.}
        \label{fig:theory}
\end{figure}
We formulate a self-consistency equation for renormalized exciton energies as
\begin{equation}
\tilde{E}_{\nu,\bQ} = E_{\nu,\bQ} + \textrm{Re}\, \Sigma(\nu,\bQ,\tilde{E}_{\nu,\bQ} / \hbar)\,,
\label{eq:Sigma_self_cons_main}
\end{equation}
with the frequency-dependent exciton self-energy
\begin{equation}
\label{eq:self-energy}
\Sigma(\nu,\bQ,\omega) = \Sigma^{\textrm{H}}(\nu,\bQ)+ \Sigma^{\textrm{F}}(\nu,\bQ)
+  \Sigma^{\textrm{PB}}(\nu,\bQ) + \Sigma^{\textrm{MW}}(\nu,\bQ,\omega). 
\end{equation}
Here, $\bQ$ denotes the exciton momentum and $\nu$ denotes the relative exciton quantum number. The unrenormalized energies $E_{\nu,\bQ}$ and corresponding exciton wave functions result from the solution of a Bethe-Salpeter equation.
Explicit expressions as well as a detailed derivation of the various self-energy contributions are given in Ref.~\cite{steinhoff2024exciton}.
The self-energy consists of Hartree (H), Fock (F), Pauli blocking (PB) and Montroll-Ward (MW) terms. 
The Hartree term contains the dipole-dipole type electrostatic interaction, responsible for the well-known exciton blueshift.
Excitons as composite bosons also experience a repulsive bosonic exchange interaction, represented by the Fock term in Eq.\,(\ref{eq:self-energy}). 
On the other hand, exchange interaction between the fermionic constituents of excitons (included in both, Hartree and Fock terms) leads to an attractive interaction as in a Fermi gas.
Finally, the Montroll-Ward self-energy contains all non-instantaneous contributions of the GW self-energy, describing frequency-dependent screening of bosonic exchange interaction by excitations within the exciton gas.
The MW term results in a decrease of the exciton energies and has been shown to largely compensate the dipolar blueshift in TMD heterobilayers \cite{steinhoff2024exciton}. Similar compensation effects have also been reported in \cite{erkensten_microscopic_2022}.

Evaluation of the self-consistency condition (\ref{eq:Sigma_self_cons_main}) at a certain temperature and density involves the calculation of exciton population functions, which we describe as Bose functions. Assuming that 
exciton-exciton interaction effects are dominated by the dipolar hIX, we distribute the exciton density only among the hIX states, with an effective temperature of $70$ K roughly corresponding to the estimated population ratio between lower and higher peak 
with the external electric field applied. The density we use for our calculations can therefore be compared to the hIE densities used in Figs.~\ref{fig:0V} and \ref{fig:1V}.

Figure \ref{fig:theory}(a) shows the predicted behavior of the A exciton depending on the exciton density for varying gate voltage. Panel (b) presents the corresponding simulations for the lower hIE. 
While the repulsive sub-meV renormalizations of the A exciton are slightly reduced with increasing electric field, the interaction between hIE becomes significantly more repulsive.
An analysis of the individual contributions to the hIE self-energy (see Appendix) shows that this is mainly due to an increase of the dipolar repulsion, whereas all other terms are much less affected by the external field. Hence, while at vanishing and small fields the net exciton-exciton potential is predicted to be attractive due to efficient excitonic screening, elevated fields lead to a dominance of the dipolar interaction over all other effects.

Our experimental zero-voltage data show similar results to those reported in earlier works \cite{datta2022highly, louca2023interspecies}, 
but quantitatively deviate from the simulations for certain conditions. This can be rooted in several reasons;
First of all, density calibration in experiment is hampered by the fact that the white-light probe introduces a constant background density, and by limitations of the rate equation model that we use to estimate the densities of individual exciton species. For example, momentum-dark exciton states, which were not included in the model, may also affect the population. 
Additionally, we observed a pronounced trion peak in pure PL measurements. Since trions carry intrinsic charge, they likely introduce additional interactions within the exciton system. Also, our simulation takes into account a quasi-thermal distribution of hIE, while the full, nonequilibrium distribution of all bright and dark exciton species is unknown. The population of excitons other than the hIE is likely to modify the exciton energy renormalization behavior. Finally, as discussed in Ref.~\cite{steinhoff2024exciton}, the GW-type exciton self-energy tends to overestimate screening effects and thus exciton redshifts.

At a gate voltage of 1 V, the A exciton exhibits very small energy shifts below 1 meV in both, experiment and simulation (for $0.1$ Vnm$^{-1}$).
The hIE, displays a pronounced blueshift in the experiment as well as theory, which is in very good qualitative agreement. Quantitatively, the blueshift in the experiment exceeds the theoretical value, which again, could be linked to the above mentioned uncertainties in the exciton density and distribution, to additional trion-exciton interactions, as well as to an overestimation of screening effects in theory. Despite the quantitative shortcomings, the significant trend of strong tunability of hIE by means of an external electric field is clearly reproduced and explained by theory.

\section{Discussion}

Our results demonstrate that hybrid interlayer excitons in bilayer MoS$_2$ exhibit strong, gate-tunable nonlinearities in the purely excitonic regime. By applying an external electric field, we have lifted the dipole degeneracy of the hIEs and enhanced their repulsive dipole-dipole interactions, which consequently dominate over attractive many-body contributions. This leads to a pronounced blueshift of hybrid excitons that by far exceeds the zero-field case. The crucial tunability is clearly explained by our microscopic theory.
In summary, we highlight that bilayer MoS$_2$ provides a platform where strong and tunable nonlinearities can be directly accessed in excitons alone. This insight is crucial for both fundamental studies of dipolar excitons and the development of future optoelectronic and polaritonic devices. 

\newpage
\textbf{Acknowledgment}:  
M.Fe. and A.S. contributed equally to this work.
S.H. acknowledges funding by the DFG via the grant Ho5194/16-1. C.S. acknowledges funding by the State of Lower Saxony, and by the German Research Foundation (DFG) within the priority programm 2244 (Project SCHN 1376/14-2). A.S. acknowledges funding by the German Research Foundation (DFG) within the priority programm 2244 (Project STE 2943/1-2) as well as computing time provided on the supercomputer Emmy/Grete at NHR@Göttingen as part of the NHR infrastructure. M.F. was supported through computational resources and services provided by Advanced Research Computing at the University of Michigan, Ann Arbor.

\textbf{Data availability}
Data available upon reasonable from the authors.

\textbf{Sample preparation}:  
The flakes for the sample were obtained by exfoliation. The stack consisting of graphene, hBN, bilayer MoS$_2$, hBN, and graphene was placed on prefabricated electrical contacts. All flakes were purchased from HQ Graphene. The contacts were partially etched into the substrate to make them flat with the rest of the surface, which facilitated the transfer of the flakes. 

\textbf{Optical measurements}:  
The measurements were done using a 532 nm green laser and a white-light laser (NKT Photonics). The white light was used to measure the reflectivity of the sample. The green laser was combined with the white light using a beam splitter, so both beams illuminated the same spot. The power of the green laser was varied while multiple measurements were averaged at each step. The objective had NA = 0.65 and 50x magnification. All measurements were taken in a liquid-helium cryostat.


\newpage


\begin{thebibliography}{25}
\providecommand{\natexlab}[1]{#1}
\providecommand{\url}[1]{\texttt{#1}}
\expandafter\ifx\csname urlstyle\endcsname\relax
  \providecommand{\doi}[1]{doi: #1}\else
  \providecommand{\doi}{doi: \begingroup \urlstyle{rm}\Url}\fi

\bibitem[Rivera et~al.(2018)Rivera, Yu, Seyler, Wilson, Yao, and
  Xu]{rivera_interlayer_2018}
Pasqual Rivera, Hongyi Yu, Kyle~L. Seyler, Nathan~P. Wilson, Wang Yao, and
  Xiaodong Xu.
\newblock Interlayer valley excitons in heterobilayers of transition metal
  dichalcogenides.
\newblock \emph{Nature Nanotechnology}, 13\penalty0 (11):\penalty0 1004--1015,
  November 2018.
\newblock ISSN 1748-3395.
\newblock \doi{10.1038/s41565-018-0193-0}.
\newblock URL \url{https://www.nature.com/articles/s41565-018-0193-0}.
\newblock Publisher: Nature Publishing Group.

\bibitem[Steinhoff et~al.(2024)Steinhoff, Wietek, Florian, Schulz, Taniguchi,
  Watanabe, Zhao, H{\"o}gele, Jahnke, and Chernikov]{steinhoff2024exciton}
Alexander Steinhoff, Edith Wietek, Matthias Florian, Tommy Schulz, Takashi
  Taniguchi, Kenji Watanabe, Shen Zhao, Alexander H{\"o}gele, Frank Jahnke, and
  Alexey Chernikov.
\newblock Exciton-exciton interactions in van der waals heterobilayers.
\newblock \emph{Physical Review X}, 14\penalty0 (3):\penalty0 031025, 2024.

\bibitem[Moody et~al.(2016)Moody, Schaibley, and Xu]{moody_exciton_2016}
Galan Moody, John Schaibley, and Xiaodong Xu.
\newblock Exciton dynamics in monolayer transition metal dichalcogenides
  [{Invited}].
\newblock \emph{JOSA B}, 33\penalty0 (7):\penalty0 C39--C49, July 2016.
\newblock ISSN 1520-8540.
\newblock \doi{10.1364/JOSAB.33.000C39}.
\newblock URL
  \url{https://opg.optica.org/josab/abstract.cfm?uri=josab-33-7-C39}.
\newblock Publisher: Optica Publishing Group.

\bibitem[Gerber et~al.(2019)Gerber, Courtade, Shree, Robert, Taniguchi,
  Watanabe, Balocchi, Renucci, Lagarde, Marie, et~al.]{gerber2019interlayer}
Iann~C Gerber, Emmanuel Courtade, Shivangi Shree, Cedric Robert, Takashi
  Taniguchi, Kenji Watanabe, Andrea Balocchi, Pierre Renucci, Delphine Lagarde,
  Xavier Marie, et~al.
\newblock Interlayer excitons in bilayer mos 2 with strong oscillator strength
  up to room temperature.
\newblock \emph{Physical Review B}, 99\penalty0 (3):\penalty0 035443, 2019.

\bibitem[Lorchat et~al.(2021)Lorchat, Selig, Katsch, Yumigeta, Tongay, Knorr,
  Schneider, and H{\"o}fling]{lorchat2021excitons}
Etienne Lorchat, Malte Selig, Florian Katsch, Kentaro Yumigeta, Sefaattin
  Tongay, Andreas Knorr, Christian Schneider, and Sven H{\"o}fling.
\newblock Excitons in bilayer mos 2 displaying a colossal electric field
  splitting and tunable magnetic response.
\newblock \emph{Physical Review Letters}, 126\penalty0 (3):\penalty0 037401,
  2021.

\bibitem[Leisgang et~al.(2020)Leisgang, Shree, Paradisanos, Sponfeldner,
  Robert, Lagarde, Balocchi, Watanabe, Taniguchi, Marie,
  et~al.]{leisgang2020giant}
Nadine Leisgang, Shivangi Shree, Ioannis Paradisanos, Lukas Sponfeldner, Cedric
  Robert, Delphine Lagarde, Andrea Balocchi, Kenji Watanabe, Takashi Taniguchi,
  Xavier Marie, et~al.
\newblock Giant stark splitting of an exciton in bilayer mos2.
\newblock \emph{Nature nanotechnology}, 15\penalty0 (11):\penalty0 901--907,
  2020.

\bibitem[Peimyoo et~al.(2021)Peimyoo, Deilmann, Withers, Escolar, Nutting,
  Taniguchi, Watanabe, Taghizadeh, Craciun, Thygesen,
  et~al.]{peimyoo2021electrical}
Namphung Peimyoo, Thorsten Deilmann, Freddie Withers, Janire Escolar, Darren
  Nutting, Takashi Taniguchi, Kenji Watanabe, Alireza Taghizadeh,
  Monica~Felicia Craciun, Kristian~Sommer Thygesen, et~al.
\newblock Electrical tuning of optically active interlayer excitons in bilayer
  mos2.
\newblock \emph{Nature Nanotechnology}, 16\penalty0 (8):\penalty0 888--893,
  2021.

\bibitem[Castellanos-Gomez et~al.(2014)Castellanos-Gomez, Buscema, Molenaar,
  Singh, Janssen, van~der Zant, and
  Steele]{castellanos-gomez_deterministic_2014}
Andres Castellanos-Gomez, Michele Buscema, Rianda Molenaar, Vibhor Singh,
  Laurens Janssen, Herre S~J van~der Zant, and Gary~A Steele.
\newblock Deterministic transfer of two-dimensional materials by all-dry
  viscoelastic stamping.
\newblock \emph{2D Materials}, 1\penalty0 (1):\penalty0 011002, April 2014.
\newblock ISSN 2053-1583.
\newblock \doi{10.1088/2053-1583/1/1/011002}.
\newblock URL \url{https://doi.org/10.1088/2053-1583/1/1/011002}.
\newblock Publisher: IOP Publishing.

\bibitem[Louca et~al.(2023)Louca, Genco, Chiavazzo, Lyons, Randerson,
  Trovatello, Claronino, Jayaprakash, Hu, Howarth,
  et~al.]{louca2023interspecies}
Charalambos Louca, Armando Genco, Salvatore Chiavazzo, Thomas~P Lyons, Sam
  Randerson, Chiara Trovatello, Peter Claronino, Rahul Jayaprakash, Xuerong Hu,
  James Howarth, et~al.
\newblock Interspecies exciton interactions lead to enhanced nonlinearity of
  dipolar excitons and polaritons in mos2 homobilayers.
\newblock \emph{Nature Communications}, 14\penalty0 (1):\penalty0 3818, 2023.

\bibitem[Erkensten et~al.(2022)Erkensten, Brem, Perea-Causín, and
  Malic]{erkensten_microscopic_2022}
Daniel Erkensten, Samuel Brem, Raül Perea-Causín, and Ermin Malic.
\newblock Microscopic origin of anomalous interlayer exciton transport in van
  der {Waals} heterostructures.
\newblock \emph{Physical Review Materials}, 6\penalty0 (9):\penalty0 094006,
  September 2022.
\newblock \doi{10.1103/PhysRevMaterials.6.094006}.
\newblock URL \url{https://link.aps.org/doi/10.1103/PhysRevMaterials.6.094006}.

\bibitem[Datta et~al.(2022)Datta, Khatoniar, Deshmukh, Thouin, Bushati,
  De~Liberato, Cohen, and Menon]{datta2022highly}
Biswajit Datta, Mandeep Khatoniar, Prathmesh Deshmukh, F{\'e}lix Thouin,
  Rezlind Bushati, Simone De~Liberato, Stephane~Kena Cohen, and Vinod~M Menon.
\newblock Highly nonlinear dipolar exciton-polaritons in bilayer mos2.
\newblock \emph{Nature communications}, 13\penalty0 (1):\penalty0 6341, 2022.

\bibitem[Wang et~al.(2018)Wang, Zhang, Liu, Li, Liu, Luo, and
  Ge]{wang2018layer}
Ting Wang, Yirui Zhang, Yuanshuang Liu, Junyi Li, Dameng Liu, Jianbin Luo, and
  Kai Ge.
\newblock Layer-number-dependent exciton recombination behaviors of mos2
  determined by fluorescence-lifetime imaging microscopy.
\newblock \emph{The Journal of Physical Chemistry C}, 122\penalty0
  (32):\penalty0 18651--18658, 2018.

\bibitem[Giannozzi et~al.(2009)Giannozzi, Baroni, Bonini, Calandra, Car,
  Cavazzoni, Ceresoli, Chiarotti, Cococcioni, Dabo, Corso, Gironcoli, Fabris,
  Fratesi, Gebauer, Gerstmann, Gougoussis, Kokalj, Lazzeri, Martin-Samos,
  Marzari, Mauri, Mazzarello, Paolini, Pasquarello, Paulatto, Sbraccia,
  Scandolo, Sclauzero, Seitsonen, Smogunov, Umari, and
  Wentzcovitch]{giannozzi_quantum_2009}
Paolo Giannozzi, Stefano Baroni, Nicola Bonini, Matteo Calandra, Roberto Car,
  Carlo Cavazzoni, Davide Ceresoli, Guido~L. Chiarotti, Matteo Cococcioni,
  Ismaila Dabo, Andrea~Dal Corso, Stefano~de Gironcoli, Stefano Fabris, Guido
  Fratesi, Ralph Gebauer, Uwe Gerstmann, Christos Gougoussis, Anton Kokalj,
  Michele Lazzeri, Layla Martin-Samos, Nicola Marzari, Francesco Mauri,
  Riccardo Mazzarello, Stefano Paolini, Alfredo Pasquarello, Lorenzo Paulatto,
  Carlo Sbraccia, Sandro Scandolo, Gabriele Sclauzero, Ari~P. Seitsonen,
  Alexander Smogunov, Paolo Umari, and Renata~M. Wentzcovitch.
\newblock {QUANTUM} {ESPRESSO}: a modular and open-source software project for
  quantum simulations of materials.
\newblock \emph{Journal of Physics: Condensed Matter}, 21\penalty0
  (39):\penalty0 395502, September 2009.
\newblock ISSN 0953-8984.
\newblock \doi{10.1088/0953-8984/21/39/395502}.
\newblock URL \url{https://doi.org/10.1088/0953-8984/21/39/395502}.

\bibitem[Giannozzi et~al.(2017)Giannozzi, Andreussi, Brumme, Bunau, Nardelli,
  Calandra, Car, Cavazzoni, Ceresoli, Cococcioni, Colonna, Carnimeo, Corso,
  Gironcoli, Delugas, DiStasio, Ferretti, Floris, Fratesi, Fugallo, Gebauer,
  Gerstmann, Giustino, Gorni, Jia, Kawamura, Ko, Kokalj, Küçükbenli,
  Lazzeri, Marsili, Marzari, Mauri, Nguyen, Nguyen, Otero-de-la Roza, Paulatto,
  Poncé, Rocca, Sabatini, Santra, Schlipf, Seitsonen, Smogunov, Timrov,
  Thonhauser, Umari, Vast, Wu, and Baroni]{giannozzi_advanced_2017}
P.~Giannozzi, O.~Andreussi, T.~Brumme, O.~Bunau, M.~Buongiorno Nardelli,
  M.~Calandra, R.~Car, C.~Cavazzoni, D.~Ceresoli, M.~Cococcioni, N.~Colonna,
  I.~Carnimeo, A.~Dal Corso, S.~de Gironcoli, P.~Delugas, R.~A. DiStasio,
  A.~Ferretti, A.~Floris, G.~Fratesi, G.~Fugallo, R.~Gebauer, U.~Gerstmann,
  F.~Giustino, T.~Gorni, J.~Jia, M.~Kawamura, H.-Y. Ko, A.~Kokalj,
  E.~Küçükbenli, M.~Lazzeri, M.~Marsili, N.~Marzari, F.~Mauri, N.~L. Nguyen,
  H.-V. Nguyen, A.~Otero-de-la Roza, L.~Paulatto, S.~Poncé, D.~Rocca,
  R.~Sabatini, B.~Santra, M.~Schlipf, A.~P. Seitsonen, A.~Smogunov, I.~Timrov,
  T.~Thonhauser, P.~Umari, N.~Vast, X.~Wu, and S.~Baroni.
\newblock Advanced capabilities for materials modelling with {Quantum}
  {ESPRESSO}.
\newblock \emph{Journal of Physics: Condensed Matter}, 29\penalty0
  (46):\penalty0 465901, October 2017.
\newblock ISSN 0953-8984.
\newblock \doi{10.1088/1361-648X/aa8f79}.
\newblock URL \url{https://doi.org/10.1088/1361-648x/aa8f79}.

\bibitem[Perdew et~al.(1996)Perdew, Burke, and
  Ernzerhof]{perdew_generalized_1996}
John~P. Perdew, Kieron Burke, and Matthias Ernzerhof.
\newblock Generalized gradient approximation made simple.
\newblock \emph{Physical Review Letters}, 77\penalty0 (18):\penalty0
  3865--3868, October 1996.
\newblock \doi{10.1103/PhysRevLett.77.3865}.
\newblock URL \url{https://link.aps.org/doi/10.1103/PhysRevLett.77.3865}.

\bibitem[Perdew et~al.(1997)Perdew, Burke, and
  Ernzerhof]{perdew_generalized_1997}
John~P. Perdew, Kieron Burke, and Matthias Ernzerhof.
\newblock Generalized {Gradient} {Approximation} {Made} {Simple}.
\newblock \emph{Physical Review Letters}, 78\penalty0 (7):\penalty0 1396--1396,
  February 1997.
\newblock \doi{10.1103/PhysRevLett.78.1396}.
\newblock URL \url{https://link.aps.org/doi/10.1103/PhysRevLett.78.1396}.

\bibitem[van Setten et~al.(2018)van Setten, Giantomassi, Bousquet, Verstraete,
  Hamann, Gonze, and Rignanese]{van_setten_pseudodojo_2018}
M.~J. van Setten, M.~Giantomassi, E.~Bousquet, M.~J. Verstraete, D.~R. Hamann,
  X.~Gonze, and G.~M. Rignanese.
\newblock The {PseudoDojo}: {Training} and grading a 85 element optimized
  norm-conserving pseudopotential table.
\newblock \emph{Computer Physics Communications}, 226:\penalty0 39--54, May
  2018.
\newblock ISSN 0010-4655.
\newblock \doi{10.1016/j.cpc.2018.01.012}.
\newblock URL
  \url{https://www.sciencedirect.com/science/article/pii/S0010465518300250}.

\bibitem[Coehoorn et~al.(1987)Coehoorn, Haas, Dijkstra, Flipse, de~Groot, and
  Wold]{coehoorn_electronic_1987}
R.~Coehoorn, C.~Haas, J.~Dijkstra, C.~J.~F. Flipse, R.~A. de~Groot, and
  A.~Wold.
\newblock Electronic structure of
  \$\{{\textbackslash}mathrm\{{MoSe}\}\}\_\{2\}\$,
  \$\{{\textbackslash}mathrm\{{MoS}\}\}\_\{2\}\$, and
  \$\{{\textbackslash}mathrm\{{WSe}\}\}\_\{2\}\$. {I}. {Band}-structure
  calculations and photoelectron spectroscopy.
\newblock \emph{Physical Review B}, 35\penalty0 (12):\penalty0 6195--6202,
  April 1987.
\newblock \doi{10.1103/PhysRevB.35.6195}.
\newblock Publisher: American Physical Society.

\bibitem[Bronsema et~al.(1986)Bronsema, De~Boer, and
  Jellinek]{bronsema_structure_1986}
K.~D. Bronsema, J.~L. De~Boer, and F.~Jellinek.
\newblock On the structure of molybdenum diselenide and disulfide.
\newblock \emph{Zeitschrift für anorganische und allgemeine Chemie},
  540\penalty0 (9-10):\penalty0 15--17, 1986.
\newblock ISSN 1521-3749.
\newblock \doi{10.1002/zaac.19865400904}.
\newblock \_eprint:
  https://onlinelibrary.wiley.com/doi/pdf/10.1002/zaac.19865400904.

\bibitem[Grimme et~al.(2010)Grimme, Antony, Ehrlich, and
  Krieg]{grimme_consistent_2010}
Stefan Grimme, Jens Antony, Stephan Ehrlich, and Helge Krieg.
\newblock A consistent and accurate ab initio parametrization of density
  functional dispersion correction ({DFT}-{D}) for the 94 elements {H}-{Pu}.
\newblock \emph{The Journal of Chemical Physics}, 132\penalty0 (15):\penalty0
  154104, April 2010.
\newblock ISSN 0021-9606.
\newblock \doi{10.1063/1.3382344}.
\newblock URL \url{https://doi.org/10.1063/1.3382344}.

\bibitem[Pizzi et~al.(2020)Pizzi, Vitale, Arita, Blügel, Freimuth, Géranton,
  Gibertini, Gresch, Johnson, Koretsune, Ibañez-Azpiroz, Lee, Lihm, Marchand,
  Marrazzo, Mokrousov, Mustafa, Nohara, Nomura, Paulatto, Poncé, Ponweiser,
  Qiao, Thöle, Tsirkin, Wierzbowska, Marzari, Vanderbilt, Souza, Mostofi, and
  Yates]{pizzi_wannier90_2020}
Giovanni Pizzi, Valerio Vitale, Ryotaro Arita, Stefan Blügel, Frank Freimuth,
  Guillaume Géranton, Marco Gibertini, Dominik Gresch, Charles Johnson,
  Takashi Koretsune, Julen Ibañez-Azpiroz, Hyungjun Lee, Jae-Mo Lihm, Daniel
  Marchand, Antimo Marrazzo, Yuriy Mokrousov, Jamal~I. Mustafa, Yoshiro Nohara,
  Yusuke Nomura, Lorenzo Paulatto, Samuel Poncé, Thomas Ponweiser, Junfeng
  Qiao, Florian Thöle, Stepan~S. Tsirkin, Ma{\textbackslash}lgorzata
  Wierzbowska, Nicola Marzari, David Vanderbilt, Ivo Souza, Arash~A. Mostofi,
  and Jonathan~R. Yates.
\newblock Wannier90 as a community code: new features and applications.
\newblock \emph{Journal of Physics: Condensed Matter}, 32\penalty0
  (16):\penalty0 165902, January 2020.
\newblock ISSN 0953-8984.
\newblock \doi{10.1088/1361-648X/ab51ff}.
\newblock Publisher: IOP Publishing.

\bibitem[Nakamura et~al.(2021)Nakamura, Yoshimoto, Nomura, Tadano, Kawamura,
  Kosugi, Yoshimi, Misawa, and Motoyama]{nakamura_respack_2021}
Kazuma Nakamura, Yoshihide Yoshimoto, Yusuke Nomura, Terumasa Tadano, Mitsuaki
  Kawamura, Taichi Kosugi, Kazuyoshi Yoshimi, Takahiro Misawa, and Yuichi
  Motoyama.
\newblock {RESPACK}: {An} ab initio tool for derivation of effective low-energy
  model of material.
\newblock \emph{Computer Physics Communications}, 261:\penalty0 107781, April
  2021.
\newblock ISSN 0010-4655.
\newblock \doi{10.1016/j.cpc.2020.107781}.
\newblock URL
  \url{https://www.sciencedirect.com/science/article/pii/S001046552030391X}.

\bibitem[Wietek et~al.(2024)Wietek, Florian, Göser, Taniguchi, Watanabe,
  Högele, Glazov, Steinhoff, and Chernikov]{wietek_nonlinear_2024}
Edith Wietek, Matthias Florian, Jonas Göser, Takashi Taniguchi, Kenji
  Watanabe, Alexander Högele, Mikhail~M. Glazov, Alexander Steinhoff, and
  Alexey Chernikov.
\newblock Nonlinear and {Negative} {Effective} {Diffusivity} of {Interlayer}
  {Excitons} in {Moir{\'{e}}}-{Free} {Heterobilayers}.
\newblock \emph{Physical Review Letters}, 132\penalty0 (1):\penalty0 016202,
  January 2024.
\newblock \doi{10.1103/PhysRevLett.132.016202}.
\newblock URL \url{https://link.aps.org/doi/10.1103/PhysRevLett.132.016202}.

\bibitem[Rösner et~al.(2015)Rösner, Şaşioğlu, Friedrich, Blügel, and
  Wehling]{rosner_wannier_2015}
M.~Rösner, E.~Şaşioğlu, C.~Friedrich, S.~Blügel, and T.~O. Wehling.
\newblock Wannier function approach to realistic {Coulomb} interactions in
  layered materials and heterostructures.
\newblock \emph{Physical Review B}, 92\penalty0 (8):\penalty0 085102, August
  2015.
\newblock \doi{10.1103/PhysRevB.92.085102}.

\bibitem[Florian et~al.(2018)Florian, Hartmann, Steinhoff, Klein, Holleitner,
  Finley, Wehling, Kaniber, and Gies]{florian_dielectric_2018}
Matthias Florian, Malte Hartmann, Alexander Steinhoff, Julian Klein,
  Alexander~W. Holleitner, Jonathan~J. Finley, Tim~O. Wehling, Michael Kaniber,
  and Christopher Gies.
\newblock The {Dielectric} {Impact} of {Layer} {Distances} on {Exciton} and
  {Trion} {Binding} {Energies} in van der {Waals} {Heterostructures}.
\newblock \emph{Nano Letters}, 18\penalty0 (4):\penalty0 2725--2732, April
  2018.
\newblock ISSN 1530-6984.
\newblock \doi{10.1021/acs.nanolett.8b00840}.
\newblock URL \url{https://doi.org/10.1021/acs.nanolett.8b00840}.

\end{thebibliography}

\newpage

\section{Appendix}

\subsection{Power to density}
To convert the measured power of the green laser to exciton densities we used the following simplified rate equations for the steady state:
\begin{equation}
    0 = P - n_0 (\alpha_B + \alpha_{hIE} + \alpha_A)
    \label{eq:pump}
\end{equation}
\begin{equation}
    0 = \alpha_B  n_B - \frac{n_B}{\tau_B} \rightarrow n_B = \alpha_B  n_0 \tau_B
\end{equation}
\begin{equation}
    0 = \alpha_{hIE}  n_{hIE} - \frac{n_{hIE}}{\tau_{hIE}} \rightarrow n_{hIE} = \alpha_{hIE}  n_0 \tau_{hIE}
\end{equation}
\begin{equation}
    0 = \alpha_A  n_A - \frac{n_A}{\tau_A} \rightarrow n_A = \alpha_A  n_0  \tau_A
\end{equation}
Here, $\alpha_i$ are the rates with which excited carriers relax from the pump reservoir into bound states, where $i=\left\lbrace A,B,hIE\right\rbrace$ label the exciton species. The pump term is calculated as $P=p/(A\Omega)$, where $p$ is the measured output power of the laser, the area $A$ is given by the laser spot size and $\Omega$ is the photon energy at 532 nm. For $B$ we also included all the states higher or at the energy of the $B$ as they showed a strong signal in PL and are expected to take up some of the excitation from the laser, thereby lowering the density of the other states.  The $\tau_i$ are the decay times of the individual exciton species, which are taken from \cite{wang2018layer} and \cite{lorchat2021excitons}. The lifetime of the hIE was estimated to be five times that of the A exciton, since it has about 20 \% the oscillator strength \cite{leisgang2020giant}. The $\alpha_i$ can be determined by taking PL measurements of the sample at different powers and fitting the counts for the three exciton species using a linear model \cite{louca2023interspecies}.
We got the following slope values for the excitons: $\alpha_B$ = 255.7, $\alpha_{hIE}$ = 39.6 and $\alpha_A$ = 200.5. These were then used to determine $n_0$ via Eq.~(\ref{eq:pump}). Once $n_0$ is established, the densities of the excitons species can be calculated. 

To calculate the densities of the low and high energy hIE the same PL measurement was performed while a field of 1V was applied. Fitting the two peaks to a Gaussian reveals a ratio of 133.3 in the low energy state to 4.1 in the high energy state.
We note that we did not take into account the density excited by the white light source for the reflectivity measurements. As the power of that source was not changed during the measurement the density due to the white light remains constant for all pump situations.

\subsection{Band Structure and Coulomb Matrix Elements}

Density functional theory (DFT) calculations for a freestanding H$^h_h$ MoS$_2$ homobilayer are carried out using QUANTUM ESPRESSO \cite{giannozzi_quantum_2009, giannozzi_advanced_2017}. We apply the generalized gradient approximation (GGA) by Perdew, Burke, and Ernzerhof (PBE) \cite{perdew_generalized_1996,perdew_generalized_1997} and use optimized norm-conserving Vanderbilt pseudopotentials~\cite{van_setten_pseudodojo_2018} at a plane-wave cutoff of $80$~Ry. Spin-orbit coupling is taken into account by performing noncollinear DFT calculations and using fully relativistic pseudopotentials. Uniform meshes (including the $\Gamma$-point) with $18\times18\times1$ k-points are combined with a Fermi-Dirac smearing of $5$~mRy. 
Using a fixed lattice constant of $a=3.16$~\AA\, \cite{coehoorn_electronic_1987,bronsema_structure_1986} and a fixed cell height of $20$~\AA, forces are minimized below $5\cdot 10^{-3}$~eV/\AA. The D3 Grimme method~\cite{grimme_consistent_2010} is used to include van-der-Waals corrections.

We use WANNIER90 \cite{pizzi_wannier90_2020} to construct a spin-resolved lattice Hamiltonian $H(\bk)$ in a 44-dimensional localized basis of Wannier orbitals (d$_{z^2}$, d$_{xz}$, d$_{yz}$, d$_{x^2-y^2}$ and d$_{xy}$ for Mo and W, respectively, p$_x$, p$_y$ and p$_z$ for S) from the DFT results. We also calculate the dielectric function as well as bare and screened Coulomb matrix elements in the localized basis without spin-orbit coupling using RESPACK \cite{nakamura_respack_2021}, assuming that the Coulomb matrix in Wannier representation is spin-independent. The values of the bare and screened Coulomb interaction were extrapolated from vacuum heights of $35$~\AA to $55$~\AA.
The external electric field is phenomenologically included by adding on-site contributions to the lattice Hamiltonian. We assume that these terms are linear in the electric field, with a slope of $\alpha_z =0.05 e$ nm \cite{lorchat2021excitons}.
By convention, all orbitals in the upper layer obtain a positive energy shift, while all orbitals in the lower layer obtain a corresponding negative shift.
Diagonalization of $H(\bk)$ yields the band structure and the Bloch states.
Screened Coulomb matrix elements in the Bloch basis are calculated as described in \cite{wietek_nonlinear_2024}. Environmental screening due to the hBN encapsulation can be taken into account according to the Wannier function continuum electrostatics approach \cite{rosner_wannier_2015}. The macroscopic dielectric function of a bilayer embedded in a vertical heterostructure is obtained by solving Poisson's equation \cite{florian_dielectric_2018}.
For most many-body calculations, we use a Brillouin zone sampling with $24\times 24\times 1$ k-points, limiting the Brillouin zone to the regions with radius $2$ nm$^{-1}$ around the K and -K points. Only the excitonic dielectric function is evaluated using a $48\times 48\times 1$ sampling. The four highest valence and four lowest conduction bands are considered. For every total exciton momentum $\bQ$, $48$ exciton eigenstates are taken into account.

\subsection{Individual Contributions to Exciton Energy Renormalizations}

Here we discuss the individual contributions to the exciton energy renormalization in the exciton self-energy, Eq.~(\ref{eq:self-energy}).
As discussed in detail in Ref.~\cite{steinhoff2024exciton}, the exciton Hartree and Fock self-energies can be further separated into direct and exchange terms, respectively:
\begin{equation}
\Sigma^{\textrm{H}}(\nu,\bQ)=\Sigma^{\textrm{H,(D)}}(\nu,\bQ)+\Sigma^{\textrm{H,(X)}}(\nu,\bQ)\, , 
\label{eq:exciton_Hartree_final}
\end{equation}
\begin{equation}
\Sigma^{\textrm{F}}(\nu,\bQ)=\Sigma^{\textrm{F,(D)}}(\nu,\bQ)+\Sigma^{\textrm{F,(X)}}(\nu,\bQ)\, .
\label{eq:exciton_Fock_final}
\end{equation}
The Pauli-blocking self-energy $\Sigma^{\textrm{PB}}(\nu,\bQ)$ and the Montroll-Ward self-energy $\Sigma^{\textrm{MW}}(\nu,\bQ,\tilde{E}_{\nu,\bQ} / \hbar)$ are evaluated as they are.
\begin{figure}[!h]
     \centering
     
          \includegraphics[width=\textwidth]{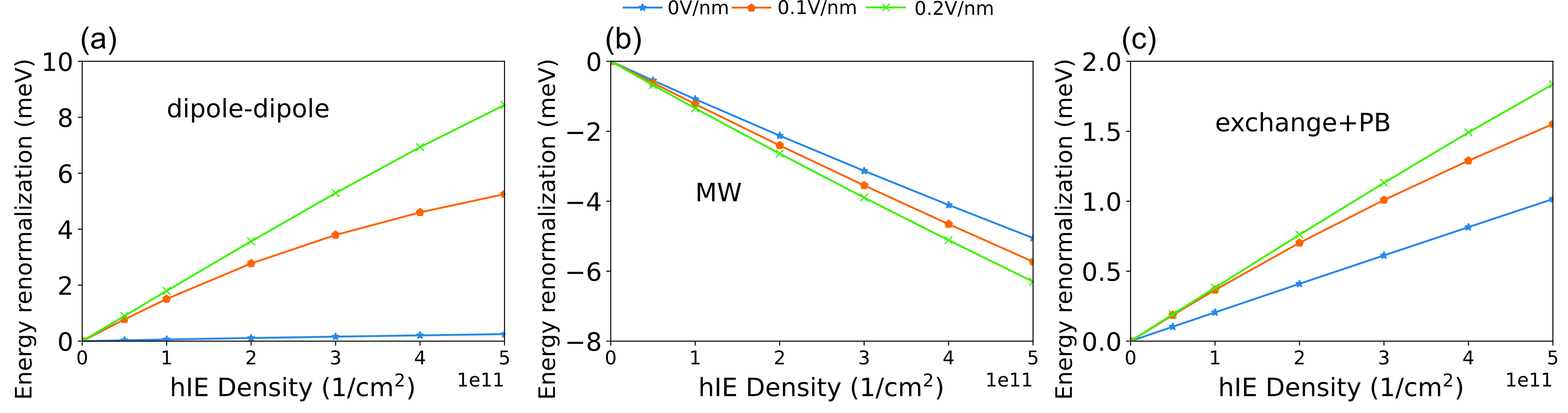}   
 
         \caption{Individual contributions to the energy renormalization of the lower hIE corresponding to the total renormalizations shown in Fig.~\ref{fig:theory}(b). The dipole-dipole contribution as direct part of the exciton Hartree self-energy is shown in (a) for different external electric fields. The Montroll-Ward contribution due to excitonic screening is shown in (b). The combined effect of fermionic exchange, contained as exchange terms in the Hartree and Fock self-energies, bosonic exchange contained as direct term in the Fock self-energy, as well as Pauli blocking is given in (c). As in Fig.~\ref{fig:theory}, the density is distributed among the different hIE according to an effective temperature of $70$ K.}
         \label{fig:theory_contrib}
 \end{figure}

Fig.~\ref{fig:theory_contrib} shows computational results for the renormalization of the lower hIE, corresponding to the total renormalizations shown in Fig.~\ref{fig:theory}(b).
We find that the energy blueshift due to dipole-dipole interaction given by $\Sigma^{\textrm{H,(D)}}$ in panel (a) is strongly enhanced by an external electric field, which polarizes the hIE and increases their inter-layer dipole moment. 
On the other hand, the energy red shift due to excitonic screening, represented by $\Sigma^{\textrm{MW}}$ in panel (b), is only weakly dependent on the electric field. The field dependence of exchange and Pauli blocking ($\Sigma^{\textrm{H,(X)}}+\Sigma^{\textrm{F}}+\Sigma^{\textrm{PB}} $) is slightly stronger, since they are directly sensitive to the population of the lower hIE; An increasing splitting of the hIE increases the quasi-thermal population of the lower branch, which leads to stronger renormalization effects. However, the magnitude of this term is by far the smallest at elevated field strength.
As a result, the dipole-dipole interaction is clearly favored by the external field, in particular compared to the Montroll-Ward screening term, which is a strong competitor in terms of the overall exciton shift behavior. 
Even though the net exciton energy renormalition might be negative at vanishing and small electric fields, the repulsive dipolar interaction will dominate at sufficiently high field strength.

\end{document}